\documentclass[astra,ms]{copernicus}

\begin{document}

\title{Possible structure in the cosmic ray electron spectrum measured by the ATIC-2 and ATIC-4 experiments}

\author[1]{A.~D.~Panov}
\author[1]{V.~I.~Zatsepin}
\author[1]{N.~V.~Sokolskaya}
\author[2]{J.~H.~Adams, Jr.}
\author[3]{H.~S.~Ahn}
\author[1]{G.~L.~Bashindzhagyan}
\author[4]{J.~Chang}
\author[2]{M.~Christl}
\author[5]{T.~G.~Guzik}
\author[5]{J.~Isbert}
\author[3]{K.~C.~Kim}
\author[1]{E.~N.~Kouznetsov}
\author[1]{M.~I.~Panasyuk}
\author[1]{E.~B.~Postnikov}
\author[3]{E.~S.~Seo}
\author[2]{J.~Watts}
\author[5]{J.~P.~Wefel}
\author[3]{J.~Wu}

\affil[1]{Skobeltsyn Institute of Nuclear Physics, Moscow State University, Moscow, Russia}
\affil[2]{Marshall Space Flight Center, Huntsville, AL, USA}
\affil[3]{University of Maryland, Institute for Physical Science \& Technology, College Park, MD, USA}
\affil[4]{Purple Mountain Observatory, Chinese Academy of Sciences, China}
\affil[5]{Louisiana State University, Department of Physics and Astronomy, Baton Rouge, LA, USA}

\runningtitle{Possible structure in the electron spectrum}

\runningauthor{A. D. Panov et al}

\correspondence{A. D. Panov (panov@dec1.sinp.msu.ru)}

\received{}
\pubdiscuss{} 
\revised{}
\accepted{}
\published{}


\firstpage{1}

\maketitle

\begin{abstract}
A strong excess in a form of a wide peak in the energy range of 300--800~GeV was discovered in the first measurements of the electron spectrum in the energy range from 20~GeV to 3~TeV  by the balloon-borne experiment ATIC \citep{CHANG-NATURE2008}. The experimental data processing and analysis of the electron spectrum with different criteria for selection of electrons, completely independent of the results reported in \citet{CHANG-NATURE2008} is employed in the present paper. The new independent analysis generally  confirms the results of \citet{CHANG-NATURE2008}, but shows that the spectrum in the region of the excess is represented by a number of narrow peaks. The measured spectrum is compared to the spectrum of \citet{CHANG-NATURE2008} and to the spectrum of the Fermi/LAT experiment .
\end{abstract}


\introduction

The ATIC (Advanced Thin Ionization Calorimeter) balloon-borne spectrometer was designed to measure the energy spectra of elements from H to Fe with an individual resolution of charges in primary cosmic rays for energy region from 50 GeV to 100 TeV. The ATIC project had three successful flights around the South Pole in 2000--2001 (ATIC-1), 2002-2003 (ATIC-2) and 2007-2008 (ATIC-4). ATIC-1 was a test flight and is not discussed below. The ATIC is comprised of a fully active bismuth germanate (BGO) calorimeter, a carbon target with embedded scintillator hodoscopes, and a silicon matrix that is used as a main charge detector. The calorimeter is comprised of 8 layers of 40 BGO crystals in each layer in ATIC-2 and of 10 layers in ATIC-4. The details of construction of the apparatus and the procedures of its calibration are given in  \citet{GUZIK2004,ZATSEPIN2004,GANEL2005,PANOV2008}. 

It has been shown that ATIC was capable of not only to measure the spectra of cosmic ray nuclear components, but also the spectrum of cosmic ray electrons plus positrons  (hereinafter referred to as the  ``electron spectrum'' for brevity). To separate the electrons from a much higher background of protons and other nuclei, the differences in the shower development in the apparatus for electrons and nuclei are used. Spectrum of electrons, measured in this way, was published in  \citet{CHANG-NATURE2008}. The most notable reported detail of the electron spectrum was an ``excess'' of electrons in the range of 300--800~GeV.  A possible connection of this ``ATIC excess'' with the nearby sources of cosmic ray electrons such as pulsars and supernova remnants or annihilation of dark matter particles, was discussed for the first time in \citet{CHANG-NATURE2008}. Such possibilities provoked a very extensive discussion in the literature.

As the result obtained in \citet{CHANG-NATURE2008} is considered to be a very important, it must be tested and confirmed in the ATIC experiment by other methods. The solution of this problem is the main objective  of this paper. The work has been carried out by the Skobeltsyn Institute of Nuclear Physics group of the ATIC collaboration starting from the low-level procedures related to the apparatus calibration and up to the analysis of the final results, completely independently on the previous works reported in \citet{CHANG2008,CHANG-NATURE2008}.

\section{Selection of electrons from input cosmic ray particle flux}

The selection of electrons from the input cosmic ray particle flux was performed using the Monte Carlo simulations of the showers generated by the primary electrons in the ATIC apparatus and in the residual atmosphere with the FLUKA system \citep{FLUKA}. To distinguish the electrons, some special quantities are constructed that describe the shape of the shower in the apparatus in the longitudinal and transverse directions in such way that these quantities take sharply different values for ``typical'' electrons and for ``typical'' protons\footnote{The nuclei with $z\ge2$ are easily rejected by the use of the charge detector.}. We call these quantities ``electron filters''. In contrast to  \citet{CHANG2008,CHANG-NATURE2008}, where only one filter was used, we constructed five different filters to provide a cross-check of the results and an evaluation of the methodological reliability. The following results address to one  filter, called Chi, but  are confirmed by other filters as well. The basic parameters for the construction of the Chi-filter are the  relative energy deposits in the layers of the calorimeter, $C_l=E_l/E$ (where $l=0,1,2,\dots$ is the number of the layer from the top to the bottom, and $E$ is the total energy deposit in the calorimeter) and the root mean square widths of the shower in the layers $R_l$. The value of the filter Chi is given by the formula
$$\mathrm{Chi}=\sqrt{\frac{1}{8}\left[\sum_{l=0}^3[(R_l-\bar R_l)/\sigma^R_l]^2 + \sum_{l=4}^7[(F_l-\bar F_l)/\sigma^F_l]^2 \right]},$$
where $F_l=R_l\sqrt{C_l}$. The mean values and dispersions of quantities $R_l$ and $F_l$ are calculated for the incident electrons by the simulations.  The distribution of  Chi-values for single-charge particles after a preliminary selection of events by energy deposits in the layers of calorimeter is shown in Fig.~\ref{FilterChiAndBack}. The preliminary energy condition consists of a number of cuts for relative energy deposits $C_l$. These cuts are the lower limits for $C_0,\dots,C_6$ and the upper limits for $C_7,C_8,C_9$. They are obtained by the simulations in which about 98\% of the electrons passed the selection with the applied cuts, while about 30\% of the protons in ATIC-2 and 70\% in ATIC-4 were rejected by these cuts. The energy cuts are based on the difference of the averaged longitudinal development of the shower in the calorimeter for electrons and protons. For the final selection of electrons, we used such cut values for  Chi that 30--25\% of the primary electrons were rejected and lost together with the rejected proton background (the optimal cut values were different for ATIC-2 and ATIC-4 flights; the efficiency of the selection was estimated by the simulations). These losses are taken into account when  evaluating  an absolute primary flux of the cosmic ray electrons.

\section{Fine structure in the electron spectrum}
\label{FINE}

The calorimeter of the ATIC spectrometer is in practice thick for electrons (it is 18 radiation lengths for ATIC-2 and 22.5 radiation lengths for ATIC-4), therefore the incident energy of an electron can be easily determined by the total energy deposit in the calorimeter. The test measurements on the electron beam at CERN \citep{GANEL2005} and the simulations have shown that the ATIC spectrometer has a very high energy resolution for the electrons. The resolution is a slow-varying function of energy and in the terms of half the line width at half maximum less than 3\% at the energies of 200--600~GeV. No unfolding procedure is needed to obtain the high-resolution spectrum with such a narrow apparatus function since the width of the apparatus function is less than or comparable with the width of the energy bins for all binnings used. Only an energy-dependent scaling factor of about 1.1, evaluated in the simulations, is used to obtain the primary energy of the electrons from the energy deposited in the calorimeter. The high value of the energy resolution enables us to investigate the electron spectrum for the presence of a structure on the scale of 0.1-0.2 decade in energy. 

It is essential that to detect a structure on the scale of 0.1-0.2 decade in energy it is not necessary to investigate the ``absolute'' electron spectrum obtained after the subtraction of the proton background (see Fig.~\ref{FilterChiAndBack}), and accounting for the electron scattering in the residual atmosphere. Neither the  background nor the scattering of electrons in the atmosphere can lead to a short-period structure in the electron spectrum. We have confirmed it by the simulations explicitly: It was shown that the atmospheric correction produces only an amplitude scaling factor that is slow-varying and almost linear function of the logarithm of energy (see Sec.~\ref{ABS} for the further details). On the other hand, the protons which can mimic the electrons ($\mathrm{Chi} \sim 1$) and are responsible for the proton background, have a very wide energy apparatus function (about 50\% of a mean deposited energy), and such an apparatus function smears all possible structures (if they are present) in the initial flux of protons. Consequently, the proton background should form a spectrum without prominent features.

The spectra of electrons as measured in  ATIC-2 and ATIC-4 experiments, without atmospheric corrections and the subtraction of proton background, are shown in Fig.~\ref{Fine} in the energy range of 30--900~GeV with the step of 0.035 decades in energy. It is easy to see a structure in the  range of 200--600~GeV, which is well reproduced in both experiments. Hereinafter, we call this phenomenon the `fine structure'. A total statistics per each energy bin of the spectrum is shown in Fig.~\ref{N-A2A4-Fine}.

The statistical significance of the observed fine structure is determined by two different factors: firstly, by the statistical significance of the presence of non-random structure with the usual $\chi^2$-criterion for the total ATIC-2 + ATIC-4 spectrum; and secondly, by the statistical significance of the correlation (similarity) of structures of the spectra measured separately in ATIC-2 and ATIC-4. The evaluation was performed for the region of the spectrum from 200~GeV to 800~GeV. The formulae for the calculation of the  $\chi^2$-value and a correlation-like function $C$ are respectively:
$$
  \chi^2 = \sum_i\left(\frac{y_i - \bar y_i}{\sigma^y_i}\right)^2;\quad
  C = \sum_i \frac{x_i-\bar x_i}{\sigma^x_i}\,\frac{y_i-\bar y_i}{\sigma^y_i},
$$
where $x_i,y_i$ are integer numbers representing the statistics in the bin number $i$; $\bar x_i,\bar y_i$ are some `smoothed' values of the spectra in the same bin; and $\sigma^x_i = \sqrt{\bar x_i},\sigma^y_i = \sqrt{\bar y_i}$ are the Poisson dispersions for the smoothed spectra; $y_i$ denotes the  intensity of the sum of ATIC-2 and ATIC-4 spectra in the formula for $\chi^2$, and $x_i,y_i$ denote the intensities of ATIC-2 and ATIC-4 spectra separately in the formula for $C$. Probabilities $P_{\chi^2} = P(\chi^2_\mathrm{rnd}>\chi^2_\mathrm{exp})$, $P_C=P(C_\mathrm{rnd}>C_\mathrm{exp})$ for the values of $\chi^2$ and $C$ of random spectra to exceed the experimental values $\chi^2_\mathrm{exp}$ and $C_\mathrm{exp}$ were calculated. The random spectra are the spectra with the Poisson statistics relative to the average values determined by smoothed spectra. Quantities $1-P_{\chi^2}$, $1-P_C$ are statistical significances of the fine structure calculated in two different ways. All probabilities are calculated with the Monte Carlo simulations and are checked approximately against the standard table values where it was possible. 

The problem of calculation of $P_{\chi^2}$, $P_C$ is no way a simple one. One difficulty of the approach is that there is some irreducible arbitrariness in the construction of a `smoothed' spectrum. Therefore, one of the main objectives was to investigate the stability of the estimate against an arbitrariness of the smoothing procedure. To obtain the smoothed spectra we approximated the primary spectra in a wide range of energies (from 35~GeV to 1500~GeV) by a cubic spline. To evaluate a stability of the statistical significance against the arbitrariness of the smoothing procedure, we studied smoothing procedures with a very wide range of steps of averaging during the construction of the spline (spline steps). We found that there exists a wide plateau in the dependence of an estimated statistical significance on the spline step. The estimated statistical significance of the fine structure is almost constant for the spline steps from 0.12 to 0.24  decades of energy. Such a plateau is possible due to a very high amplitude of the fine structure of the spectrum. The details of any 'reasonable' smoothing procedure are not actually important. Our final results are obtained by averaging with spline steps 0.12, 0.15, 0.18, 0.21, and 0.24. 

The other difficulty is that the estimate is highly dependent on the specific energy binning used in the building of the spectrum. The significance for each spline step (see above) is actually evaluated by averaging on a number of different binning with bin widths from 0.015 to 0.035 decades of energy. The errors of the final estimates are calculated as standard deviations representing the fluctuations of the estimates with different binning and with different `splining'. The contribution to the error from an instability related to the bin size is dominant.

In this way we found that the statistical significance for the correlation is $1-P_C=(99.69^{+0.25}_{-1.32})\%$, and for the $\chi^2$-criterion $1-P_{\chi_2}=(99.68^{+0.27}_{-1.69})\%$. The high statistical significance practically eliminates a random nature of the observed fine structure. We would like to note that the present estimates of the statistical significance should be considered as preliminary ones. It is desirable to develop a `binningless' technique to avoid the relatively large statistical errors of the estimates caused by the fluctuations of the estimates against binning. This work is currently in progress.

Several tests were also performed to exclude possible methodological causes of the observed structures. The statistics of the proton background in the range of the filter values near the electron peak but free of the electron signal ($2<\mathrm{Chi}<3.5$, see Fig.~\ref{FilterChiAndBack}) were investigated: no signs of a possible structure were found. The different electron filters were studied: all the filters produced a very similar structure. The spectra for different solid angles and different time periods of the experiments were compared: the fine structure was reproduced in all the cases. Thus, no evidence that the observed fine structure might be caused by the methodological effect was found.

If the existence of the fine structure in the electron spectrum is confirmed by independent experiments, then its most likely source will be the nearby supernova remnants and/or pulsars, but not the annihilation or decay of the dark matter particles. The idea that the cosmic ray electron spectrum in the TeV energy range may have features related to a few nearby sources like the pulsars is forty years old \citep{SHEN1970}. The possibility that the deviation from a smooth power-like behavior in the spectrum is caused by the individual nearby sources and can  be observable in this energy range  was proposed in \citet{NISHIMURA1980}, in relation to the feasible experiments. Later on this issue was considered from various viewpoints repeatedly (see, for example, \citet{POHL1998,EW2002,PROFUMO2008}). Specifically, a structure very similar to the one observed in the ATIC experiment is predicted in \citet{MALYSHEV2009}, where it is especially emphasized that such a structure might be used as a signature to distinguish between the annihilation or decay of dark matter particles and the other sources of electrons, such as nearby pulsars. The dark matter can not be a source of a specific structure represented by several narrow peaks \citep{MALYSHEV2009}. The reason is simple and deep: the sources like the pulsars can be treated as instantaneous, meanwhile the sources like the dark matter halos or subhalos (clumps) should be considered as permanent in time.  High energy wing of the spectrum of an instantaneous source can, in principle, create a very sharp peak in the electron spectrum, due to the process of cooling of electrons \citep{MALYSHEV2009}, therefore a number of sources can produce several peaks as observed in the ATIC experiment. On the contrary, the permanent sources like the dark matter clumps would mix such peaks with different energies related to the different moments of the time of emission of electrons and produce wide distributions in the energy spectrum \citep{KUHLEN2009} which has little in common with the fine structure observed by the ATIC. Even if some `multi-peak' structure were produced by the dark matter clumps, this structure would be very smooth and smeared compared to the observed ATIC fine structure (see \citet[FIG.3]{BRUN2009}, \citet[FIG.9]{CLINE2010}).

\section{The electron spectrum after proton background subtraction and atmospheric correction}
\label{ABS}

The spectrum shown in Fig.~\ref{Fine} does not provide a basis for a comparison with the results of other experiments, since it does not give a correct absolute intensity of the electron flux. To obtain the correct absolute intensity of the electron spectrum, the proton background must be subtracted, and the electron spectrum must be corrected for the scattering in the residual atmosphere (and we neglect the secondary atmospheric electron background  from a hadronic component of cosmic rays, since it is shown to be small  at the thickness of a residual atmosphere of about 5~g/cm$^2$ typical for the ATIC flights \citep{NISHIMURA1980}).

\begin{figure}[t]
\vspace*{2mm}
\begin{center}
\includegraphics[width=8.3cm]{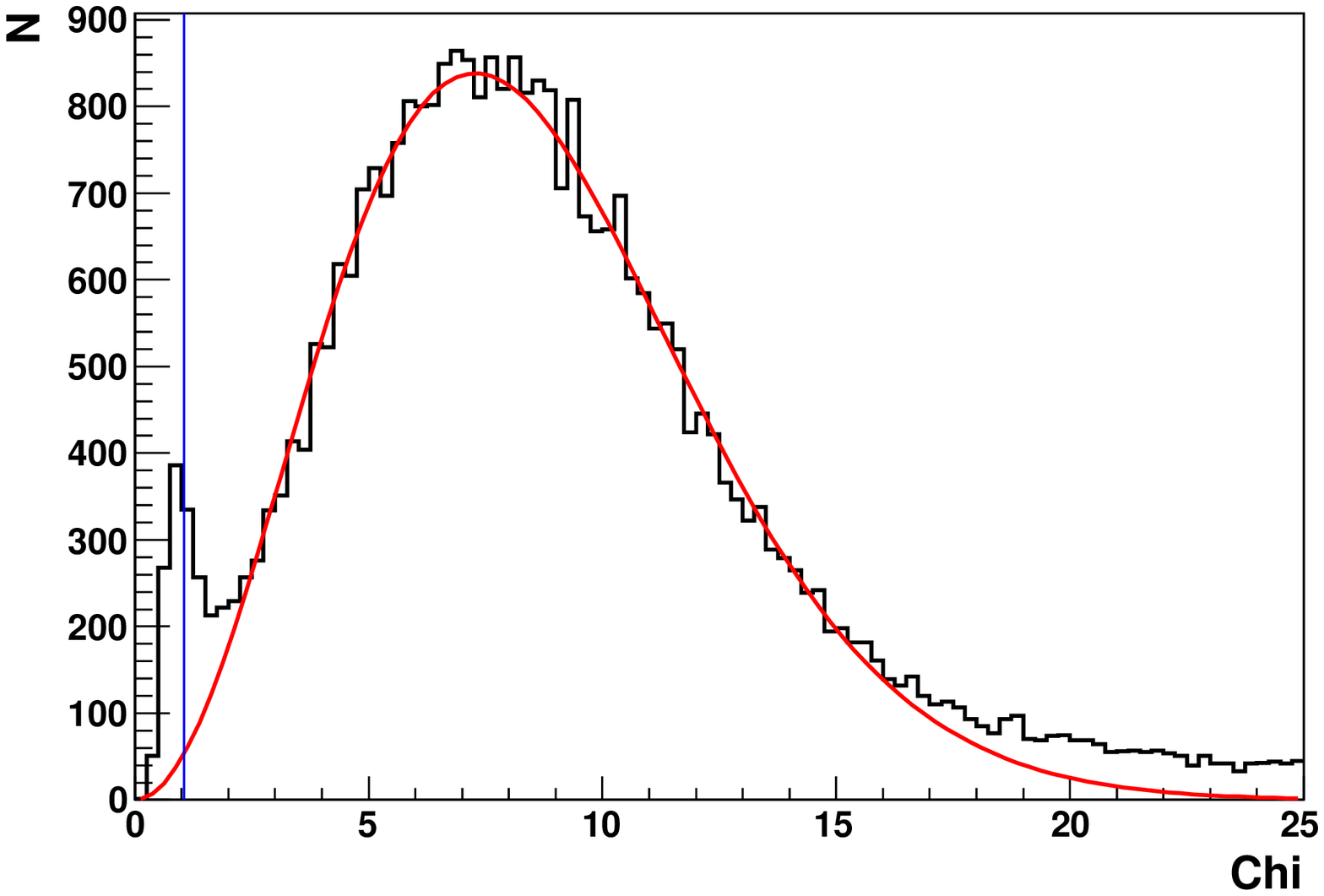}
\end{center}
\caption{Black solid line -- the distribution of the filter Chi values for single-charge particles (ATIC-2, the energy deposit range  100--200~GeV). The narrow peak on the left side is related to electrons, the wide peak is related to protons. Blue line represents the cut value to select electron events. The cut value was optimized for the best signal to background relation. Red line shows the approximation of the proton distribution for  subtracting of proton background (see Section~\protect\ref{ABS}). The approximation presented in this picture is from three-parameter assemblage $y(x) = A x^2 \exp[-(x/\sigma)^\alpha]$, the region to fit is from $\mathrm{Chi}=2$ to $\mathrm{Chi}=16$.} 
\label{FilterChiAndBack}
\end{figure}

\begin{figure}[t]
\vspace*{2mm}
\begin{center}
\includegraphics[width=8.3cm]{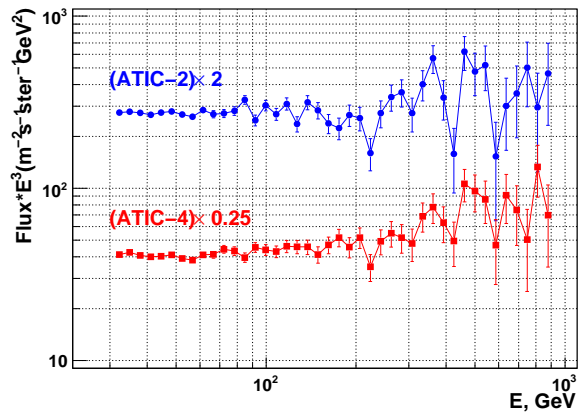}
\end{center}
\caption{The spectrum of electrons without the subtraction of  proton background and without an atmospheric correction
as measured in the ATIC-2 and ATIC-4 experiments.} 
\label{Fine}
\end{figure}

\begin{figure}[t]
\vspace*{2mm}
\begin{center}
\includegraphics[width=8.3cm]{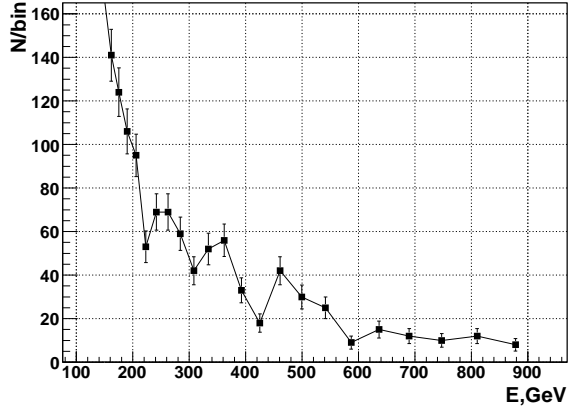}
\end{center}
\caption{Total statistics of the electron spectrum in ATIC-2 $+$ ATIC-4 flights in each energy bin. The number of events in the range of 200--810~GeV, studied for statistical significance of the structure, is 701.} 
\label{N-A2A4-Fine}
\end{figure}

The procedure of the subtraction of proton background based on the simulations of proton cascades in the ATIC apparatus leads to an unstable result: the inevitable small errors in the simulations lead to large errors in the electron spectrum \citep{PANOV2009}. In this paper, we implemented a method that is independent of the simulations. The method is based on an approximation of the experimental plot  of Chi-values for protons (see Fig.~\ref{FilterChiAndBack}) with the simple functions and on the interpolation of these functions to the coordinates origin. We examined the functions from two different three-prpameter assemblages $y(x) = A x^2 \exp[-(x/\sigma)^\alpha]$ and $y(x) = Ax/[1+(x/\sigma)^\alpha]$, which produce a reasonable approximation of the proton peak, but result in somewhat different extrapolations of the proton background under the electron peak. The scale of the possible systematic error related to the background subtraction is estimated by comparison of the results for different types of functions. This method is not rigorous, of course, and should be considered as a qualitative estimate. The scale of the proton background is about 5\% of the total amplitude of the measured spectrum near 40~GeV and about 40\% near 700~GeV. The total estimated corridor of possible systematic errors varies from ${}^{+15\%}_{-16\%}$ for the energy of 40~GeV to ${}^{+57\%}_{-46\%}$ at 700~GeV, and is related mainly to the uncertainty in the detection efficiency at low energies, while at high energies it is related to the errors in the subtraction of backgrounds. The systematic errors can not lead to an essential distortion in the shape  of the spectrum.

\begin{figure}[t]
\vspace*{2mm}
\begin{center}
\includegraphics[width=8.3cm]{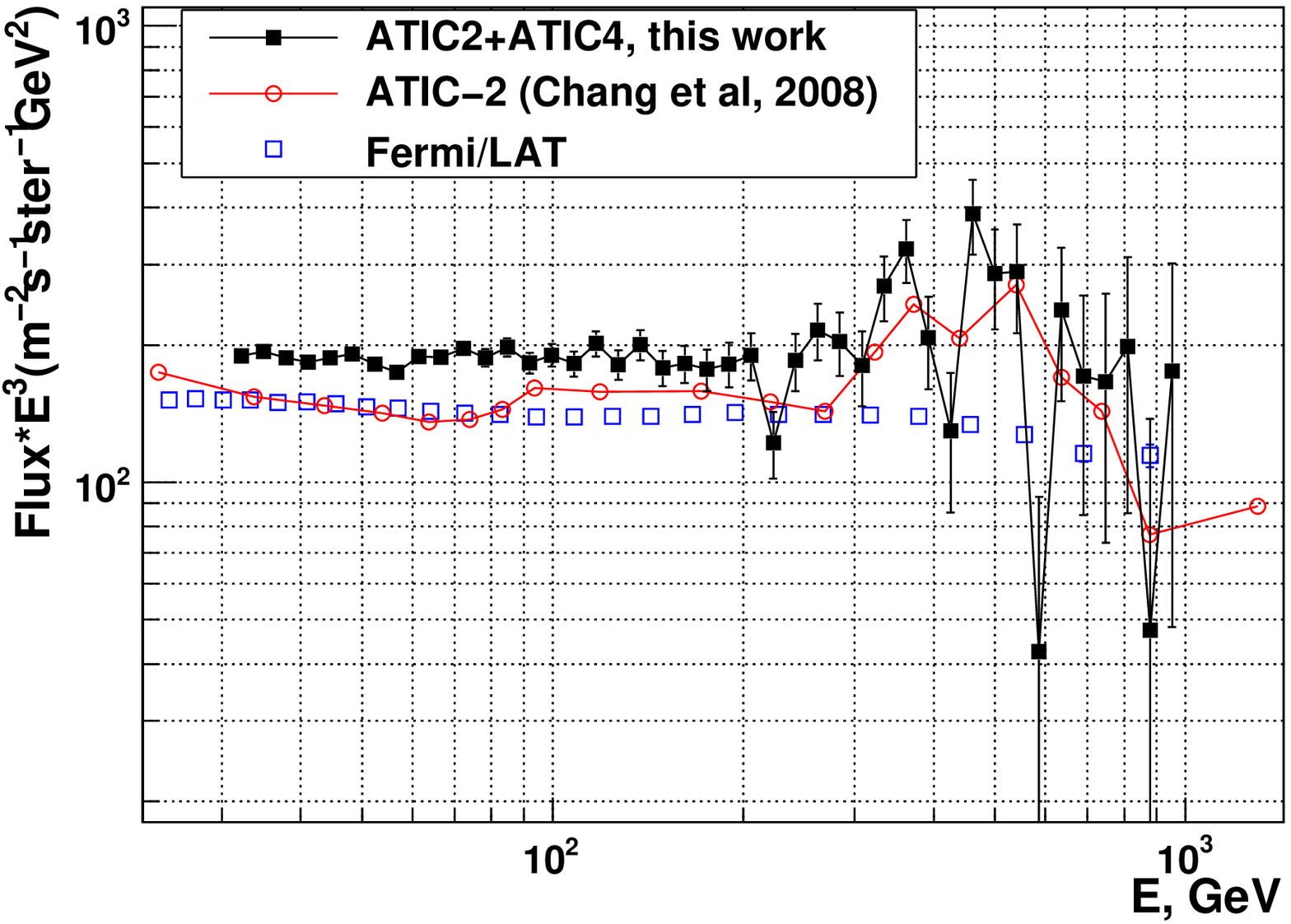}
\end{center}
\caption{ The total \mbox{ATIC-2 + ATIC-4} electron spectrum of this work after subtracting the proton background and after an atmospheric correction, the ATIC spectrum of the paper \citet{CHANG-NATURE2008} and the results of Fermi/LAT \citep{FERMILAT2009}. To simplify the picture, the experimental errors in ATIC spectrum of \citet{CHANG-NATURE2008} are not shown. The estimated corridor of possible systematic errors for the ATIC spectrum of the present work varies from ${}^{+15\%}_{-16\%}$  at 40~GeV to ${}^{+57\%}_{-46\%}$ at 700~GeV.} 
\label{ToAbs}
\end{figure}

An atmospheric correction is calculated on the basis of the simulations of the scattering of primary electrons in the atmosphere using the FLUKA system \citep{FLUKA}. As suggested in our paper \citet{PANOV2009}, one does not need to correct the energy at the top of the ATIC apparatus to obtain the energy at the top of the atmosphere since the scattering angles of the secondary gamma quanta are very small, and the energy of an electron is recorded in the calorimeter together with the energies of almost all secondary gamma quanta. However, these gamma quanta may distort the shape of the cascade in the apparatus, which leads to some additional inefficiencies of the filtration of the electrons. This was taken into account. The correction factor calculated from the simulations is approximated by a linear function of the logarithm of energy, and varies from 1.42 at 30~GeV to 1.26 at 700~GeV, forcing the spectrum to become more steep. Absolute electron spectrum measured in the present paper, along with the ATIC electron spectrum of the paper \citet{CHANG-NATURE2008} and the  results of the space spectrometer Fermi/LAT \citep{FERMILAT2009} are shown in Fig.~\ref{ToAbs}.

\conclusions

Our results confirm the existence of the ATIC excess, but this excess is resolved into a fine structure. For the energies below 200 GeV, our spectrum is identical in form to the spectrum of Fermi/LAT. The fine structure measured by ATIC above 200 GeV can be smoothed by using energy bins as wide as in the Fermi/LAT spectrum. Moreover, the ATIC spectrum is related to the part of the South sky with declination between $-45^{\mathrm o}$ and $-90^{\mathrm o}$, while the Fermi/LAT spectrum integrates the flux over the whole sky. Taking into account the possible anisotropy of the electron energy spectrum, it would be more correct to compare the ATIC spectrum with the spectrum measured by Fermi/LAT for the same part of the sky and with narrower energy bins. 
 
 \begin{acknowledgements}
The work is supported by grant of RFBR 08-02-00238.
 \end{acknowledgements}




%



\addtocounter{figure}{-1}\renewcommand{\thefigure}{\arabic{figure}a}

\end{document}